\documentclass[aps,prl,twocolumn,showpacs,amssymb,amsmath]{revtex4}
\usepackage{graphics,epsfig,subfigure,graphicx}

\begin{document}

\title{Geographical Coarsegraining of Complex Networks}
\author{Beom Jun Kim}
\affiliation{Department of Molecular Science and Technology, Ajou University,
Suwon 442-749, Korea}

\begin{abstract}

We perform the renormalization-group-like numerical analysis
of geographically embedded complex networks on the two-dimensional
square lattice.
At each step of coarsegraining procedure, the four vertices 
on each $2 \times 2$ square box are merged to a single vertex, 
resulting in the coarsegrained system of the smaller sizes.
Repetition of the process leads to the observation that the 
coarsegraining procedure does not alter the qualitative characteristics 
of the original scale-free network, which opens the possibility of
subtracting a smaller network from the original network
without destroying the important structural properties.
The implication of the result is also suggested in the context
of the recent study of the human brain functional network.

\end{abstract}


\maketitle

Study of complex networks has been one of the most
active research areas not only in physics but also
in other various disciplines of natural and social sciences~\cite{ref:network,WS,BA}.
In some existing networks, the computerized automatic
data acquisition techniques make it possible to grab the
detailed information of interconnections in networks.
In contrast, in many  biological and social networks,
the complete network structure is hard to be defined and
even when it is possible it requires tremendous
time-demanding efforts.
The detailed structure of the neuronal network of 
{\it Caenorhabditis elegans}~\cite{WS}, composed of about 300 neuron 
cells and 14 synaptic couplings per neuron, has been obtained 
by biologists through direct observations.
In comparison to the {\it C. elegans} neural network, the complexity
of the human brain is gigantic: It contains about $10^{11}$ neuron cells, 
each of which is connected to $10^3$-$10^4$ other neurons via synaptic
couplings. Construction of the detailed map of all neuron connections in
human brain is beyond imagination and will be so in the future.

In the viewpoint of statistical physics, on the other hand,
understanding the qualitative collective behavior of the brain, 
although it originates from the actual detailed interactions of 
neurons and abundant biochemical substances, may not require such
detailed microscopic map of interneuron connections.
In this regard, the recent study by Egu{\'i}luz {\it et al.} in 
Ref.~\onlinecite{brain} draws much interest: Brain activity has been
measured from $32 \times 64 \times 64$ sites (called voxels)
and the intervoxel correlation has been used to map out the
functional network of the human brain. Although the number of voxels
in Ref.~\onlinecite{brain} is more than a million, each voxel
still contains $O(10^5)$ neuron cells. Consequently, one can say
that the brain functional network in Ref.~\onlinecite{brain} has been 
based on {\it heavily coarsegrained} information, and thus it is
not clear whether the observed scale-freeness of the network is 
a genuine emerging property of actual interneuron connections or not.
Very recently it has been found that a very simple vertex merging
process results in scale-free networks, {\it regardless
of the initial network structure}~\cite{merging}. 
In the present context, this observation may suggest that if the network is 
too much coarsegrained, one cannot trust the resulting scale-free
distribution since it may not reflect the structure of the
original network but is a simple artifact of coarsegraining.

In this Letter, we start from model scale-free networks that are geographically
embedded~\cite{rozenfeld}, and then repeat several steps of geographic 
coarsegraining.
We find that the coarsegraining process does not change important
properties of the original network. In particular, the degree exponent
$\gamma$, the clustering property, the assortative feature, and the hierarchical structure
do not change much upon the iteration of the geographic coarsegraining.
Our result suggests that the scale-free feature of the human brain
functional network may not be the artifact of the coarsegraining,
and thus the increase (or the decrease) of the size of voxels 
 is expected not to change the main
results of Ref.~\onlinecite{brain}. We also suggest that one can
use the geographic coarsegraining method presented in this Letter
to subtract a smaller network from the original larger network,
without destroying important structural properties. This can
be very useful when the network is too big to be handled 
for a given computational capability. 

We first build the geographically embedded scale-free network following
Ref.~\onlinecite{rozenfeld}: $N = L \times L$ vertices are put on
lattice points of the two-dimensional square net, and then the degree $k$
of each vertex is chosen according to the degree distribution function
$p(k) \propto  k^{-\gamma}$. A vertex $v$ is selected
at random and then its assigned degree $k_v$ is realized  on the basis that
the geographically closer vertices 
(within the distance proportional to $\sqrt{k_v}$) are connected first.
As the procedure is repeated over vertices, some vertices may not fulfill 
their assigned degrees if their all possible target vertices already exhausted
their allowed number of edges. When this happens, those vertices have
degrees different from the initially assigned ones, and there appears  
the cutoff degree scale (and the corresponding cutoff length scale) 
beyond which $p(k)$ deviates from the power-law form $p(k) \sim k^{-\gamma}$
(see Ref.~\onlinecite{rozenfeld} for details). 

\begin{figure}
\centering{\resizebox*{0.3\textwidth}{!}{\includegraphics{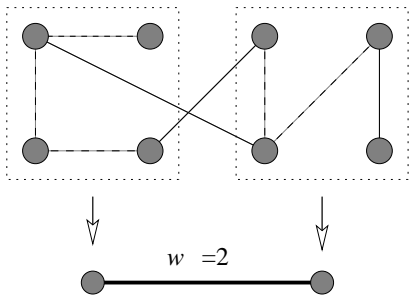}}}
\caption{Coarsegraining procedure. Four vertices in each $2\times 2$
box is merged to a single vertex. After merging, 
two edges (thin solid lines) connecting 
vertices in different boxes becomes one edge (thick solid line)
with the weight $w=2$. This procedure is repeated for all 
$2\times 2$ square boxes resulting in the coarsegrained network 
which is four times smaller than the original network. We then
remove edges of smaller weights in order to fix the average degree.
}
\label{fig:procedure}
\end{figure}

Once the network is constructed
in this way, we repeat the following geographic coarsegraining, which
is in parallel to the Kadanoff block spin renormalization group procedure
in standard statistical mechanical systems (see Fig.~\ref{fig:procedure}):
Four vertices on each square box of the size $2 \times 2$ is merged to a single
vertex and accordingly the edges connecting intra-box vertices (dashed lines 
in Fig.~\ref{fig:procedure})
are disregarded, but the inter-box connections are kept (thin solid line
in Fig.~\ref{fig:procedure}). We also keep
track of how strong the edges are by assigning the weight $w_{vw}$ that
is simply the number of edges connecting two merged vertices $v$ and $w$.
For example, in Fig.~\ref{fig:procedure} there exist two edges (thin solid
line) connecting the two square boxes before the merging, which gives rise
to the weight $w=2$ for the edge (thick solid line) connecting the two 
vertices after the merging. 

If we keep all the edges in the coarsegrained
network, the average degree increases as the procedure is iterated,
resulting in the fully-connected network eventually. To remedy this,
we fix the average degree at each step of coarsegraining by removing
weaker edges with smaller values of the weight. Suppose that 
we have to remove $M_r$ edges to keep the average degree the same,
and that there are $M_w$ edges of the weight $w$. For example, 
for $M_r < M_{w=1}$, randomly picked $M_r$ edges of $w=1$ is removed. 
If $M_{w=1} < M_r < M_{w=2}$, all $w=1$ edges removed and $M_r - M_{w=1}$
edges with $w=2$ are randomly deleted .
The above procedure makes sense since in real situations, it is common
that the coarsegraining is often accompanied by the change of
the sensitivity of the measurement: When the system is looked at
a far distance, we only have interest in large scale structures.

\begin{figure}
\centering{\resizebox*{0.48\textwidth}{!}{\includegraphics{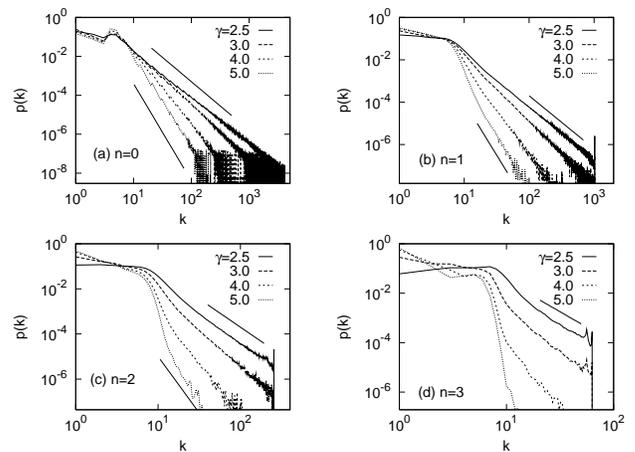}}}
\caption{Degree distribution $p(k)$ versus the degree $k$ for
geographically embedded scale-free networks. 
The original $64 \times 64$ networks in (a) with the degree exponents 
$\gamma=2.5, 3.0, 4.0$, and 5.0 are coarsegrained $n$ times;
(b) $n=1$, (c) $n=2$, and (d) $n=3$.
Clearly shown is that the degree exponent does not change upon
the iteration of coarsegraining. Full lines in (a)-(c) are for the power 
law distributions with the exponents $2.5$ and $5.0$,
while in (d) only the line for the exponent $2.5$ is shown for comparison.
$\gamma$ values in (b)-(d) only indicate the degree exponents
for the corresponding initial networks.
}
\label{fig:all}
\end{figure}

\begin{figure}
\centering{\resizebox*{0.48\textwidth}{!}{\includegraphics{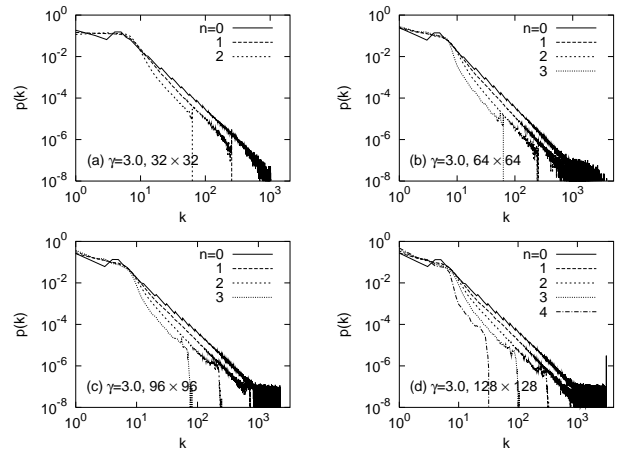}}}
\caption{Degree distribution $p(k)$ versus the degree $k$ 
for the networks with the degree exponent $\gamma = 3$.
The original network sizes are (a) $32 \times 32$, (b) $64 \times 64$,
(c) $96 \times 96$, and (d) $128 \times 128$.
The resulting coarsegrained network at the $n$th iteration
displays the same degree exponent.
}
\label{fig:allN}
\end{figure}

Figure~\ref{fig:all} shows the result of the coarsegraining.
Original networks of the size $64 \times 64$  are generated following
Ref.~\onlinecite{rozenfeld} for $\gamma=2.5, 3.0, 4.0, 5.0$ and
the above explained coarsegraining process is iterated. 
The network size in this work is smaller than the cutoff length
scale beyond which the network ceased to be scale free, which
is also seen in Fig.~\ref{fig:all}(a) where the cutoff degree
scale is absent (see Ref.~\onlinecite{rozenfeld}). 
One sees clearly that the coarsegraining process does not change the
degree exponent $\gamma$. In the terminology of the renormalization
group (RG) formalism, the scale-free network with any value of $\gamma$
is the stable fixed point of the RG flow. This observation implies
that the scale-free network in Ref.~\onlinecite{rozenfeld} possesses
neither the degree scales nor the length scales upto the cutoff
length scale which is larger than the network size in the present
study. 
In Fig.~\ref{fig:all}, as the coarsegraining process is iterated
the scale-free region appears only for sufficiently large $k$ regions.

We then study finite-size effects in Fig.~\ref{fig:allN}, which
shows $p(k)$ at the $n$-th iterations for initial networks
of various sizes (a) $32 \times 32$, (b) $64 \times 64$, (c) $96 \times 96$, 
and (d) $128 \times 128$ (all for $\gamma=3$).
Clearly exhibited is that as the initial network size is increased the network
still remains to be scale-free even after many steps of iterations, which then 
excludes the possibility that observed behaviors are finite-size artifacts.
The scale-free degree distribution detected in the human brain {\it functional}
network~\cite{brain} does not actually imply that the neural network
of human brain is scale-free. One reason is because the technique
in Ref.~\onlinecite{brain} only measures the functionality correlation
of two separate voxels, not the actual path of voxels through which
biochemical signal transfers. One can also argue that since each
voxel contains large number of neurons [about $O(10^5)$], the
observed scale-free distributions can be the artifact of the
coarsegraining, considering the recent study in Ref.~\onlinecite{merging}
that scale-free distributions can emerge from merging.
Our main results~\cite{foot} in the present study implies that this is not
the case and that accordingly the scale-free distribution in human brain
functional network is expected to be the genuine property
of the brain, not the artifact of the coarsegrained information.

\begin{figure}
\centering{\resizebox*{0.48\textwidth}{!}{\includegraphics{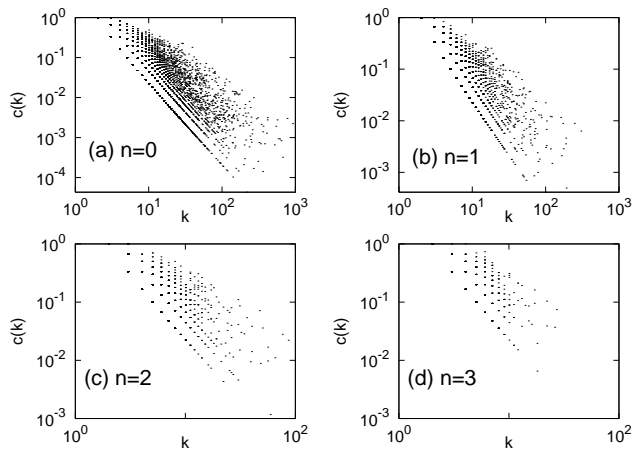}}}
\caption{The clustering coefficient $C(k)$ versus $k$ for the $128\times 128$
network with $\gamma = 3$ at the $n$-th iteration step of coarsegraining
for $n=$ (a) 0 , (b) 1, (c) 2, and (d) 3. The qualitative feature 
remains the same upon coarsegraining procedure.
}
\label{fig:ck}
\end{figure}
\begin{figure}
\centering{\resizebox*{0.48\textwidth}{!}{\includegraphics{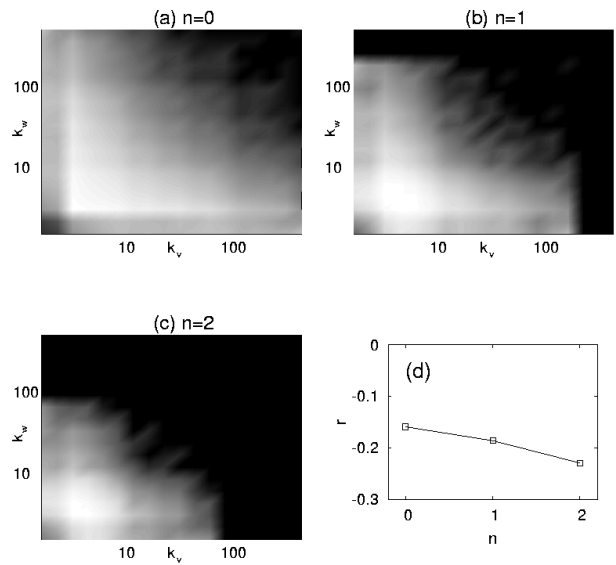}}}
\caption{(a)-(c) Density plot of the degrees $k_v$ and $k_w$ where
$v$ and $w$ are two vertices connected by each edge. Brighter region
indicates that there are more edges in that region. The initial
network shows disassortative behavior (higher $k_v$ prefers lower $k_w$,
and vice versa), which remains qualitatively the same as the 
coarsegraining procedure proceeds $n$ times: $n=$ (a) 0 (initial network),
(b) 1, and (c) 2. (d) Assortativity coefficient $r$ versus the number $n$
of iterations of coarsegraining. After the coarsegraining, the network
remains to be disassortative.
}
\label{fig:kk}
\end{figure}

We next investigate other important structural properties
of networks. Many real networks including Internet, World Wide Web,
and the actor network, are characterized by the existence of 
the hierarchical structure~\cite{ravasz,ala}, which can be
usually detected by the negative correlation between 
the clustering coefficient(see Ref.~\onlinecite{WS}) and the 
degree~\cite{ravasz}.  
For example, the  Barab{\'a}si-Albert network~\cite{BA},
which does not possess  hierarchical structure, is known to have the 
clustering coefficient $C_v$ of the vertex $v$ 
independent of its degree $k_v$ 
[i.e., $C(k) \sim k^0$, see Ref.~\onlinecite{ravasz}], while
Holme-Kim model~\cite{HK} has been shown to have 
$C(k) \sim k^{-1}$~\cite{szabo}, in accord with the observations of 
many real networks~\cite{ravasz}. In Fig.~\ref{fig:ck}, we plot $C(k)$ at
the $n$-th iteration step of the coarsegraining for the initial
network of the size $128 \times 128$ with $\gamma = 3$. The
geographically embedded network in Ref.~\onlinecite{rozenfeld}
is found to be somehow special since $C(k)$ is better described
by $C(k) \sim k^{-2}$ rather than the abundantly found $C(k) \sim
k^{-1}$. But this feature remains the same upon the coarsegraining
procedure, implying that the coarsegraining does not change the
hierarchical structure of the network.

We next study the assortative mixing 
characteristics~\cite{newman,maslov} of the network.
For the assortative network, vertices with the higher degree
tend to have high-degree neighbor vertices, while for the disassortative
network, higher degree vertices favor to have lower degree neighbors.
The degrees $k_v$ and $k_w$ of the two vertices $v$ and $w$
connecting each edge is measured and then the histogram is computed
by using $20 \times 20$ bins in log-log scales in $k_v$-$k_w$ plane.
The brightness of the region in Fig.~\ref{fig:kk} (a)-(c) is 
chosen in proportion to the logarithm of the height of the histogram
in that region. Again found is that the coarsegraining procedure 
does not change the disassortative mixing property of the network,
i.e., at any iteration step, the high-degree vertices in the
network tend to have low-degree neighbors.
This behavior of the disassortative mixing can also be detected by
the assortativity  coefficient $r$ (see Ref.~\onlinecite{newman}
for the definition). If $r$ has positive value, the network 
has assortative mixing property while it is disassortative otherwise.
In Fig.~\ref{fig:kk}(d), $r$ is shown to have negative values at
the $n=0,1,2$ coarsegraining steps. The decrease of $r$ with $n$
is not completely understood, although this dependence
of $r$ versus the network size $N$ appears to be consistent with 
Ref.~\onlinecite{newman}, where $r$ tends to approach zero from
below as the larger disassortative network is considered.

So far we have introduced a geographical coarsegraining
procedure and applied it to  the geographically embedded scale-free
network in Ref.~\onlinecite{rozenfeld}. Although the network sizes
become smaller as the coarsegraining procedure proceeds, it has
been found that several key features of the initial networks 
do not change qualitatively. In particular, the degree exponent $\gamma$
does not change, and hierarchical structure [detected by  the
negative correlation between the clustering coefficient $C(k)$ 
versus degree $k$] and the disassortativity (detected by that
more edges connect high-degree vertices to low-degree vertices
than to high-degree vertices) are remain qualitatively the same.
Our geographic coarsegraining procedure can be useful when
the initial network is of a huge size since one can then
systematically reduce the network size without destroying
important characteristics of the network. Modification of the
present coarsegraining method to apply for the network which
is not geographical embedded can be an interesting extension.
The main results also suggest that the scale-free distribution
found recently for the human brain functional network may not be
the artifact due to the large voxel size (each contains $O(10^5)$
neuron cells), but the genuine property of the brain.

\begin{figure}
\centering{\resizebox*{0.48\textwidth}{!}{\includegraphics{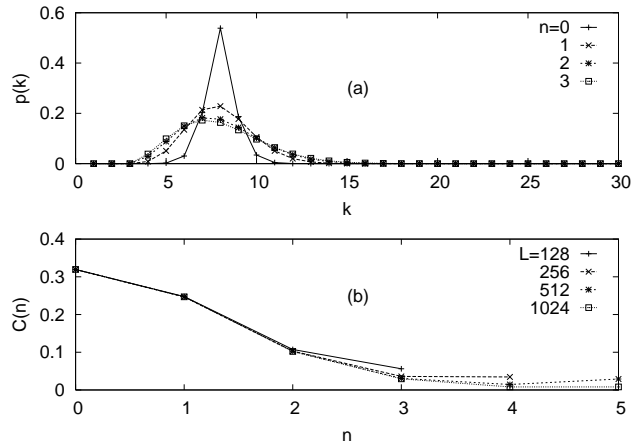}}}
\caption{Geographically embedded WS network at the
rewiring probability $P=0.1$. (a) Degree distribution $p(k)$ 
   at the iteration steps $n=0$, 1, 2, and 3. (b) Clustering
   coefficient $C(n)$ versus $n$.
}
\label{fig:ws}
\end{figure}

We finally study the two-dimensional Watts-Strogatz (WS) network, built
similarly to Ref.~\onlinecite{WS}: Vertices are put on
the two-dimensional square lattice points and every vertex is
connected to its nearest  and next-nearest neighbor vertices.
Each edge is visited once, and with the rewiring probability $P$,
is rewired to a randomly chosen vertex. The resulting network
belongs to the so-called exponential network since the
tail in the degree distribution is exponentially small.
We then iterate our geographic coarsegraining procedure with
the average degree kept constant at each iteration.
In Fig.~\ref{fig:ws}(a), the initial network of the size $128 \times 128$
at the rewiring probability $P=0.1$ is coarsegrained
$n$ times. As $n$ becomes larger, the degree distribution  remains
to be exponential and tends to saturate. 
In Fig.~\ref{fig:ws}(b), initial two-dimensional WS networks of 
various sizes $L=128, 256, 512$ and 1024 at $P=0.1$ are coarsegrained 
and the clustering coefficient $C(n)$ is plotted as a function of 
the number $n$ of iterations. As the coarsegraining process
proceeds the clustering coefficient is shown to decrease towards zero,
which indicates that the RG stable fixed point of the WS network
is close to the random network of Erd\"os and R\'enyi.

\acknowledgments
The author acknowledges H. Jeong and J.D. Noh for useful suggestions.
This work has been supported by the Korea Science
and Engineering Foundation through Grant
No. R14-2002-062-01000-0.
Numerical computations have been
performed on the computer cluster Iceberg at Ajou University.


\begin{thebibliography}{30}

\bibitem{ref:network}For general reviews, see, e.g.,
D.J. Watts, {\em Small Worlds} (Princeton University Press, Princeton, 1999);
R. Albert and A.-L. Barab{\'a}si, Rev. Mod. Phys. {\bf 74}, 47 (2002);
S.N. Dorogovtsev and J.F.F. Mendes, {\it Evolution of Networks}
(Oxford University Press, New York, 2003).

\bibitem{WS}
D.J. Watts and S.H. Strogatz, Nature (London) {\bf 393}, 440 (1998).

\bibitem{BA}
A.-L. Barab{\'a}si and R. Albert, Science {\bf 286},  509  (1999);
A.-L. Barab{\'a}si, R. Albert, and H. Jeong, Physica A {\bf 272},  173  (1999).

\bibitem{brain} V.M. Egu{\'i}luz, D.R. Chialvo, G. Cecchi, M. Baliki, and
A.V. Apkarian, cond-mat/0309092.

\bibitem{merging} B.J. Kim, A. Trusina, P. Minnhagen, and K. Sneppen, nlin.AO/0403006.

\bibitem{rozenfeld} A.F. Rozenfeld, R. Cohen, D. ben-Avraham, and S. Havlin,
Phys. Rev. Lett. {\bf 89}, 218701 (2002); D. ben-Avraham, A.F. Rozenfeld, R. Cohen, and S. Havlin, Physica A {\bf 330}, 107 (2003).

\bibitem{foot} The main results in this work are obtained
for the networks embedded in two dimensions while the human brain
network is embedded in three dimension. The extension of the present
study to a higher dimension can be interesting.

\bibitem{ravasz} E. Ravasz and A.-L. Barab{\'a}si, Phys. Rev. E {\bf 67},
026112 (2003). 

\bibitem{HK} P. Holme and B.J. Kim, Phys. Rev. E {\bf 65}, 026107 (2002).

\bibitem{szabo} G. Szab{\'o}, M. Alava, and J. Kert{\'e}sz, Phys. Rev. E {\bf 67}, 056102 (2003).

\bibitem{ala} A. Trusina, S. Maslov, P. Minnhagen, and K. Sneppen, Phys. Rev.
Lett. {\bf 92}  178702 (2004).


\bibitem{newman} M.E.J. Newman, Phys. Rev. Lett. {\bf 89}, 208701 (2002);
M.E.J. Newman and J. Park, Phys. Rev. E {\bf 68}, 036122 (2003).

\bibitem{maslov} S. Maslov and K. Sneppen, Science {\bf 296}, 910 (2002).


\end{thebibliography}
\end{document}